\documentclass{emulateapj}

\usepackage{apjfonts}
\shorttitle{Sgr A* in context}
\shortauthors{S. Markoff}

\begin{document}

\title{Sgr A* in context: daily flares as a probe of the
  fundamental X-ray emission process in accreting black holes}
\author{Sera Markoff\altaffilmark{1}} \affil{Massachusetts Institute
  of Technology, Center for Space Research, Rm. NE80-6035, Cambridge,
  MA 02139} \email{sera@space.mit.edu}

\altaffiltext{1}{NSF Astronomy \& Astrophysics
Postdoctoral Fellow}

\begin{abstract}

Our central Galactic supermassive black hole, Sgr A*, exists mostly in
a very stable, extremely low-luminosity ($\sim 10^{-9} L_{\rm Edd}$),
thermal quiescent state, which is interrupted roughly daily by a
brief, nonthermal X-ray flare.  Because they are not accompanied by
significant changes in the radio wavelengths, the flares make Sgr A*
unusual in the context of black holes accreting at slightly higher
rates.  Those sources display a radio/X-ray luminosity correlation
whose normalization scales with central mass, and that holds over
orders of magnitude in accretion power.  There is significant scatter
in this correlation, due in part to measurement uncertainties and
intrinsic variability.  By studying the correlation in sources
bracketing Sgr A* in radio luminosity and whose physical parameters
are well measured, we can derive a statistical measure of this local
scatter.  We find that Sgr A* in quiescence and the lower intensity
flares fall well below the correlation in X-ray luminosity.  The
brightest flares are consistent within the scatter, which may indicate
an upper bound on the X-ray luminosity.  This trend is suggestive of a
state transition at the extreme low end of accretion activity, only
above which the radio/X-ray correlation is tracked.  This scenario is
easily testable because it must fulfill three unique observational
predictions: 1) As long as Sgr A* remains at its current radio
luminosity, no X-ray flare will be seen which statistically exceeds
the prediction of the correlation, 2) no source already on the
correlation will be seen to flare in the X-rays similar to Sgr A*
(i.e., without corresponding increases in the radio luminosity), and
3) sources below a critical accretion rate or luminosity will show
similar flares as Sgr A*, on timescales appropriate to their masses.

\end{abstract}


\section{Correlations in Sgr A* vs. Other Black Holes}

The anomalously low luminosity of Sgr A* ($\sim 10^{-9} L_{\rm Edd}$;
  \citealt{MeliaFalcke2001}) has puzzled researchers for over two
  decades, raising questions about its relationship to other, more
  typical, active nuclei.  It seems unlikely that Sgr A* is the only
  one of its kind; if it simply represents the lowest end of the
  luminosity scale, its behavior should map onto trends we detect in
  other accreting black hole sources.  In this Letter, we propose
  three observationally verifiable predictions to probe Sgr A*'s
  relationship to more canonical black hole sources.

Sgr A*'s proximity (8 kpc; \citealt{Reid1993,Eisenhaueretal2003}) has
resulted in the constraining of its physical parameters better than
almost any other galactic nucleus, with the exception of NGC 4258
\citep[e.g.,][]{Herrnstein1997}.  Studies of Sgr A*'s orbiting central
cluster stars reveal a $4\times10^6M_\odot$ mass
\citep{Schoedeletal2003,Ghezetal2003b}, which along with the similarly
well-constrained distance, allows us to assess Sgr A*'s relationship
to other sources with known parameters.

Despite Sgr A*'s extremely weak high-energy activity, its radio
characteristics are typical of other low-luminosity AGN (LLAGN), M81*
in particular
\citep[e.g.,][]{Ho1999,Brunthaleretal2001,Boweretal2002}.  Its steady
X-ray spectrum ($L_{\rm X}\sim2\times10^{33}$ erg/s) is soft
($\Gamma\sim2.7$) and extended, arguing for a thermal origin
\citep{Baganoffetal2003}.  Recent theoretical models developed to
explain its behavior have focused variously on inflow scenarios
\citep[e.g.,][]{LiuMelia2002,YuanQuataertNarayan2003}, outflow
scenarios \citep{FalckeMarkoff2000} and combinations of the two
\citep{YuanMarkoffFalcke2002}.  Most of these models have been applied
to other low-luminosity sources with some success, indicating that Sgr
A* shares many characteristics with its brighter cousins, but may just
represent the most underluminous extreme.

Sgr A* showed the first signs of AGN-like activity in the second cycle
of {\em Chandra} observations, with a dramatic ($\sim$50x increase)
nonthermal, hard flare on timescales of tens of minutes
\citep[$\Gamma\sim1.3$;][]{Baganoffetal2001}.  Further observations
have established that flaring occurs about once a day, with typical
increases of 5--10x in flux \citep{Baganoff2003}.  The cm-radio
emission, however, has not yet been seen to vary by more than a factor
of a few \citep{MeliaFalcke2001}.

While the multiwavelength variability characteristics have not yet
been fully determined, the ``submm bump'' in Sgr A*'s spectrum, which
includes the IR band, is clearly related to the flaring X-ray
component \citep[e.g.][]{Eckartetal2004}.  The physical origin is
still being debated \citep[see articles in, e.g.,][]{GC2002}, however
most quiescent-state models for Sgr A* can be adapted to explain the
flares.  A consensus has formed that the submm/IR variability is due
to synchrotron emission from mildly relativistic, quasi-thermal
electrons very close to the central object, while the X-ray flares are
due to either a continuation of this synchrotron emission due
to a hard tail in the distribution, synchrotron self-Comptonized
emission (SSC) or combinations of the two
\citep[e.g.,][]{Markoffetal2001,LiuMelia2002,YuanQuataertNarayan2004}.
These magnetic mechanisms dominate in what seems to be the absence of
a canonical \citep{ShakuraSunyaev1973} thin disk
\citep{FalckeMelia1997,LiuMeyerHofmeister2004}.  

Often low-luminosity, accreting X-ray binaries (XRBs) in their
low/hard state (LHS; see \citealt{McClintockRemillard2003}) are
associated with compact jets and share a general morphology with
LLAGN.  A correlation between the radio and X-ray emission was first
detected in the Galactic XRB, GX 339-4 \citep{Hannikainenetal1998}.
Later observations showed that this correlation holds over a few
orders of magnitude changes in source luminosity with time
\citep{Corbeletal2000, Corbeletal2003}.  The same universal
relationship now appears to apply to all LHS XRBs with comparable
broadband data \citep{GalloFenderPooley2003}.

Scale-invariant jet synchrotron models predict this correlation
\citep{Markoffetal2003} as a consequence of how their radio emission
scales with power \citep{FalckeBiermann1995}, and the correlation can
be expressed via the dependence of the X-ray emission on the accretion
rate \citep{HeinzSunyaev2003}. These scalings led to two independent
proposals of a unification between accreting black holes from stellar
to galactic scales
\citep{MerloniHeinzDiMatteo2003,FalckeKoerdingMarkoff2004}.  When the
observed X-ray fluxes of various AGN samples are scaled to compare
with XRB-mass black holes, they fall roughly on the same correlation
defined by a single LHS XRB as it evolves in time.  There is
significant scatter, however, which may result from uncertainties in
the measured physical parameters, beaming effects, and/or intrinsic
variations.  The overall success of this formulation, however,
supports a further unification of certain black hole sources in terms
of their accretion power, as well as their orientation.


\begin{figure*}
\centerline{\hbox{\includegraphics[width=.5\textwidth]{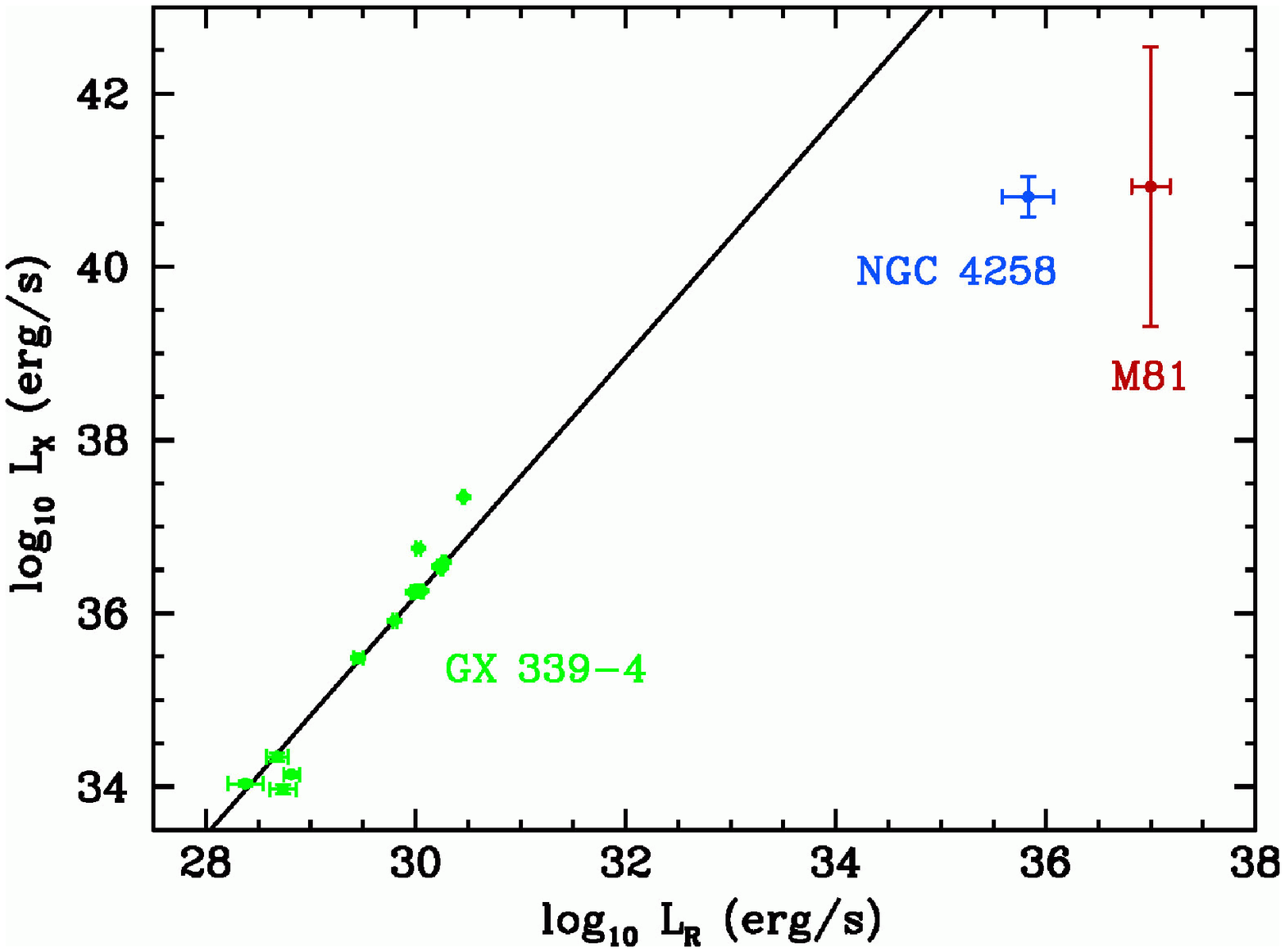}\hspace*{.13in}\includegraphics[width=.49\textwidth]{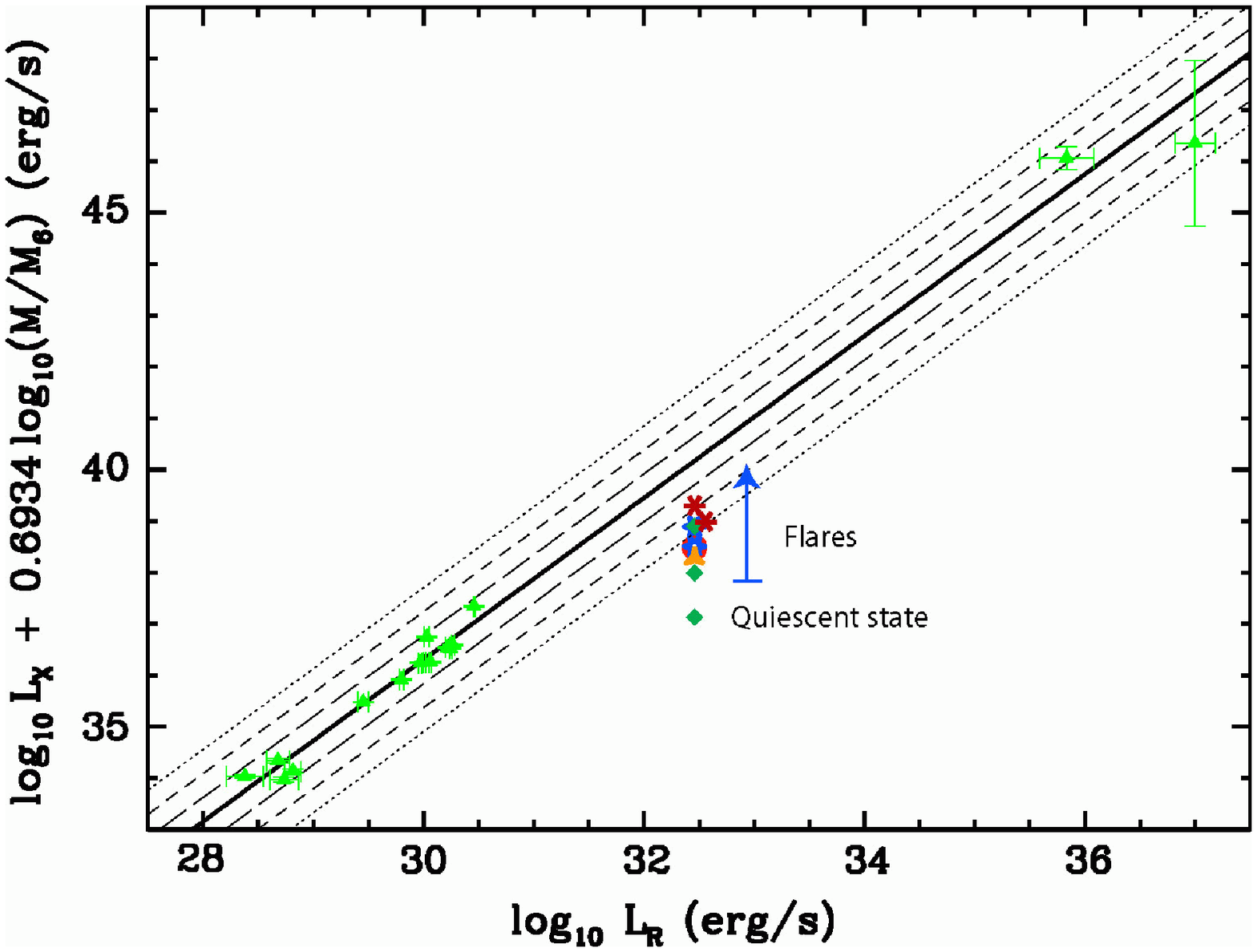}}}
\caption{a) Radio luminosity (at 5 GHz) vs. 3--9 keV integrated X-ray
luminosity for the three well-measured sources bracketing Sgr A* along
the radio/X-ray correlation: the X-ray binary GX 339$-$4 (ATCA and
{\em RXTE} data from \citealt{Corbeletal2003}, reanalyzed using new
response matrices by \citealt{Nowaketal2004}), and the LLAGN NGC~4258
and M81.  The latter two sources are represented by average
luminosities with standard deviations based on the best available
observations (see text).  The solid line is the jet synchrotron
model-predicted correlation based on the GX~339$-$4 data
\citep{Markoffetal2003}.  The sources do not fall along the same
correlation because the X-ray luminosities have not yet been scaled
for mass.  b) The derived local mass-scaled radio/X-ray correlation.
The solid line indicates the average value for the correlation
coefficients after Monte Carlo simulations, $\log_{10} L_{\rm X} =
-(10.275\pm1.982) + (1.575\pm0.067)\log_{10}L_{\rm
R}-(0.692\pm0.080)\log_{10}(M/6M_\odot)$, with contours in the average
scatter $<$$\sigma$$>$ from the correlation represented as
increasingly finer dashed lines. All observations of Sgr A*, both in
quiescence and flaring states, are included.  It is clear that the
quiescent state and weak flares are not consistent with the
correlation, while the larger flares are within the scatter.
\label{xradio}}
\end{figure*}  


\section{Statistical Analysis}

In order to make a statistical statement about Sgr A*'s relationship
to more typical low-luminosity black holes, we consider the region of
the correlation space around Sgr A*.  A detailed statistical analysis
for the overall correlation was already conducted by
\cite{MerloniHeinzDiMatteo2003}, but this included a large
sample of AGN for which there are often large uncertainties in the
physical parameters.  The authors thus used a symmetric linear
regression technique which attempts to compensate for the measurement
uncertainties.  In contrast, we will here focus on the radio/X-ray
correlation as defined by only a few well-constrained sources.
Sgr A* is bracketed by the multiple observations of GX~339$-$4, where
the correlation was discovered, as well as by observations of the
nearby LLAGN NGC~4258 and M81*.  Because these sources have
well-defined physical parameters and are not thought to be highly
beamed, we can hope to make a reasonable statement about the
correlation and its intrinsic scatter.  We use a Bayesian analysis: we
assume an intrinsic Gaussian scatter about the correlation, defined
excluding Sgr A*, and then assess the probability that the Sgr A* data
are consistent with this assumption.  For the three sources, intrinsic
scatter about the correlation far outweighs the measurement errors.

For our bracketing sample we include all simultaneous GX~339$-$4
observations from \cite{Corbeletal2003}, the X-ray portion of
which has been reanalyzed using the newest detector response matrices
compared to the earlier papers. For NGC~4258 and M81 we tabulate the
measured radio and X-ray luminosities from the last 20 years
\citep[][and
refs. therein]{Herrnsteinetal1999,BowerFalckeMellon2002,Pageetal2003,YoungWilson2004}
in order to define a mean and standard deviation for both.  The data
are shown in Fig.~\ref{xradio}a.  Based on the results of
\cite{MerloniHeinzDiMatteo2003} and \cite{FalckeKoerdingMarkoff2004},
we assume a form for the X-ray luminosity ${\rm log}_{10}L_{\rm X} =
C_0 + C_1{\rm log}_{10} L_{\rm R} + C_2 {\rm log}_{10}(M/M_\odot)$,
where $L_{\rm X}$ and $L_{\rm R}$ are the X-ray and radio luminosities
in erg/s, and $M$ is the black hole mass. We assume that $M$ is known,
and choose to study the measured $L_{\rm X}$ as determining the
intrinsic scatter in the correlation (as is appropriate for the
hypothesis that $L_{\rm X}$ is underluminous in Sgr A* for a known
$L_{\rm R}$).  Because we are looking at the comparison of Sgr
A*'s ${L_{\rm X}}$ as a function of $M$ and $L_{\rm R}$, we use a
nonsymmetric linear regression routine (i.e., which is not appropriate
for studies of the correlation itself as conducted by, e.g.,
\citealt{MerloniHeinzDiMatteo2003}).
 
For a given linear regression, we derive values of $\log_{10} L_{\rm
X,lr}$ for comparison with the measured values, $\log_{10} L_{\rm
X,i}$, which we assume to be Gaussian-distributed about the
correlation.  We then apply the Bayes theorem to determine the
distribution for the variance associated with this scatter,
$P(\sigma)\propto{\sigma}^{-1} \Pi_i \exp [-( \log_{10} L_{\rm X, lr}
- \log_{10} L_{\rm X, i})^2/2\sigma^2]$.  To incorporate our
uncertainty as to the best linear regression values for the
correlation, we use the Monte Carlo technique by assuming Gaussian
error distributions (determined from our tabulated data), and generate
$10^4$ samples of ($L_{\rm X}$,$L_{\rm R}$,$M$).  Linear regression
then yields $10^4$ values for the correlation coefficients, $C_{0-3}$.
For each run, we calculate a normalized probability distribution for
$P(\sigma)$, and then average these normalized probability
distributions over all runs.  This yields an average value for the
intrinsic scatter in the data about the linear correlation,
$<$$\sigma$$>$.  Our results are shown in Fig.~\ref{xradio}b, where we
plot the data with $L_{\rm X}$ scaled by the factor
$<$$C_2$$>$$\log_{10}(M/6M_\odot)$, to compare with GX~339$-$4.  The solid
line shows the average correlation with contour lines for
1--3$<$$\sigma$$>$.  While our value for $C_1$ is the same, we find a
different value for the mass scaling $C_2$ compared to
\cite{MerloniHeinzDiMatteo2003,FalckeKoerdingMarkoff2004}.  This is
most likely due to a combination of the reanalyzed GX~339$-$4 data,
and the dominant measurement uncertainties from their larger AGN
samples.  

During quiescence, Sgr A* lies $\gtrsim$ 6$<$$\sigma$$>$ below the
correlation, while the flares span $\sim$1.5--5$<$$\sigma$$>$.  We
plot all flare observations to date, from both {\em Chandra}
\citep[filled diamonds and triangle;][ and
priv.comm.]{Baganoffetal2001,Baganoffetal2003,Baganoff2003} and {\em
XMM-Newton} \citep[circle and asterisks;][
respectively]{Goldwurmetal2003,Porquetetal2003}.  There is quite a
discrepancy in spectral index between the brightest flares seen from
the two missions.  We have therefore also included stars representing
the two brightest flares after reanalysis with a different dust model
by \cite{TanDraine2004}.  Depending on which dust analysis is correct,
the highest flares seen to date fall within $\sim$1--3$<$$\sigma$$>$
of the correlation, while the lowest flares deviate significantly.  No
flare has yet statistically exceeded the prediction of the
correlation, and this trend suggests that the flares may saturate at
this upper bound.  In the next section we suggest three predictions
for observations in the coming decades which will help clarify Sgr
A*'s relationship to other, more typical accreting black holes who
track the correlation.

\section{Testable Predictions}

Based on the observations, we suggest that {\em the fundamental
  radio/X-ray correlation defines an upper limit to the X-ray flux in
  Sgr A*'s flares}.  For this to hold true, it would mean that the
  process responsible for the flares either saturates or undergoes a
  state change once a critical accretion power is reached, and
  afterwards tracks the correlation.  There are three necessary
  predictions of these scenarios, all testable within the next
  decades:

\begin{enumerate}
\item {\bf As long as Sgr A* remains at its relatively steady radio
emission level, no X-ray flares will be detected which statistically
exceed the prediction of the radio/X-ray correlation}.  If we consider
3$<$$\sigma$$>$ a significant deviation, then we would not expect any
flares with an integrated 3--9 keV luminosity exceeding an unscaled
value of $\sim 3.7\times10^{37}$ erg/s.  This is only a factor of
$\sim100$ times the brightest flare seen so far, if the dust analysis
of \cite{Porquetetal2003} is correct.  If the correlation represents a
state transition occurring above a certain accretion rate, then a
flare of this magnitude would be accompanied (with a time lag
appropriate to plasma propagation times along the jet) by an increase
in radio luminosity, to $L_{\rm R}\sim 2\times 10^{33}$ erg/s, a
factor of about 8--10.  An increase of greater than a few in the cm
band has never been observed over the last two decades of VLA
monitoring of Sgr A* \citep{Boweretal2002}, arguing that such
occurrences would be very rare.  Flares on the order of factor 1000x
over quiescence would test the limits of the correlation while
falling within past radio observation limits.
  
\item {\bf No sources already on the radio/X-ray correlation will be
  seen to flare in the X-rays without corresponding radio increases to
  keep them on the correlation.}  Several nearby LLAGN present
  themselves as sources to monitor for possible flaring, M81* in
  particular because of its similarities to Sgr A* in morphology and
  radio emission properties.  In fact, this source will be monitored
  in 2005 with {\em Chandra} for 300ks.  

\item {\bf Black holes accreting near Sgr A*'s accretion rate, in
  Eddington units, should fall below the correlation and/or show
  similar flares}.  Such sources, whether galactic or stellar scaled,
  are currently very hard to detect.  The lowest luminosity (in
  Eddington units) quiescent Galactic black hole observed to date is
  V404 Cygni, which has been studied in radio and X-ray wavelengths
  down to $\sim10^{-6} L_{\rm Edd}$ \citep{GalloFenderPooley2003}.
  Sgr A* has so far never achieved more than $10^{-8} L_{\rm Edd}$
  during flares and we know its accretion rate is $ < 10^{-6}
  \dot{m}_{\rm Edd}$ (assuming $10\%$ efficiency) in quiescence
  \citep{Boweretal2003}.  Thus it seems that if there is a critical
  transition, it occurs between $\dot{m}\approx 10^{-8}$--$10^{-6}
  \dot{m}_{\rm Edd}$.  To probe this power and below for other sources
  will likely require the use of planned X-ray missions with more
  collecting area, such as {\em Constellation-X} and {\em XEUS}, as
  well as higher sensitivity radio arrays such as the {\em EVLA} and
  {\em SKA}.  One XRB in particular, A 0620$-$00, is an ideal
  candidate since it is a brighter quiescent source, and it will also
  be monitored in 2005 (Gallo, priv. comm.).  
\end{enumerate}

The results of these observational tests
will clarify Sgr A*'s relationship to other accreting sources.  If the
predictions are not confirmed, Sgr A* must be significantly different
compared to all other accreting black holes which define the
correlation.  This would have serious consequences for theoretical
models of Sgr A* which invoke the same processes as other weak
galactic nuclei.  If the predictions are confirmed, it will help
determine which X-ray emission process dominates at the minimum level
of accretion activity.

While several accretion models tend to favor scenarios in which a thin
disk exists down to quiescent accretion levels, Sgr A* does not show
any sign of such a mode.  Most of the sources which fall on the
correlation, however, do show signs of a thin disk: e.g. reflection
and fluorescent line features, and even occasionally masers (e.g. NGC
4258; \citealt{Herrnstein1997}).  If the brightest flares never track
the correlation, the implied accretion rate in Sgr A* can be used as a
constraint on models of thin disk formation.   

Sources which follow the radio/X-ray correlation likely share the same
emission mechanisms.  Because the mechanisms behind Sgr A*'s flares
are well determined, the flares' relationship to the correlation can
thus be studied for clues about the correlation-generating processes.
In LHS XRBs, as well as in AGN, the ``standard model'' for the hard
X-ray emission involves a corona of thermal electrons which upscatters
seed photons from the underlying accretion disk
\citep[e.g.,][]{HaardtMaraschi1991}.  But at hard X-ray luminosities
already consistent within scatter to the correlation prediction, Sgr
A* still shows no signs of a thin disk mode for its accretion.  Even
if a thin disk begins to form at the presumed transition luminosity
where Sgr A* starts to track the correlation, the weak thermal photons
would not be able to completely dominate the submm-bump synchrotron
photon field for Compton upscattering.  The ability of thermal photons
to take precedence would depend on the accretion rate, and the
geometry of the scattering region.  If the upscattering plasma is
beamed away from the disk \citep[see,
e.g.,][]{Beloborodov1999,MarkoffNowak2004}, the contribution from
thermal photons would be reduced.  We suggest that if Sgr A* is shown
to either saturate at, or track, the radio/X-ray correlation during
the brightest flares, this would be a strong argument for
synchrotron-related processes (including SSC) as the fundamental
high-energy dissipative process in weakly accreting black holes, only
later supplemented at higher luminosities by thermal processes.

\section{Conclusions}

Current observations are suggestive of an upper bound on the X-ray
flares in Sgr A*, provided by the fundamental radio/X-ray correlation.
We present three necessary observational predictions, testable in the
next several years, which will probe the nature of Sgr A*'s
relationship, if any, to the correlation.  If Sgr A*'s flares saturate
at or even track the correlation at the highest luminosities, we would
argue that the responsible synchrotron-related mechanisms are then
also the dominant X-ray emission processes contributing to the
correlation at low luminosities.  Since there is no evidence for a
thin disk signature in Sgr A*, and since the brightest flares seen so
far already seem consistent with the correlation, the accretion rate
at which a disk can form steadily must be above what we have seen so
far in Sgr A*.  Once a disk forms, thermal photons would increasingly
contribute to the photon pool for Comptonization with higher accretion
rates.  If the synchrotron/SSC plasma is at all beamed, however,
thermal photons would not compete with the rest frame photons at
current luminosities.

Models which are consistent with this picture are those which can
explain the flares in Sgr A*, including nonthermally enhanced
radiatively inefficient accretion flows
\citep[RIAFs;][]{YuanQuataertNarayan2003} and/or nonthermal processes
at the base of the jets \citep{Markoffetal2001}.  Geometrically thick
accretion flows have already been discussed as a preferential
launching point for jets \citep{Meier2001}, as well as magnetic
coronae \citep[e.g.,][]{MerloniFabian2002}, although coronal formation
in the absence of a thin disk would still need to be worked out.  In
reality, there may be only semantical differences between RIAFs,
coronae and jet bases, an idea we have begun to explore elsewhere
\citep[][, Markoff, Nowak \& Wilms, in prep.]{MarkoffNowak2004}.  The
outcome of the observational tests proposed here will place limits on
the role of thermal vs. nonthermal processes, the necessary conditions
for thin disk formation and the relationship between inflow and
outflow in the weakest accreting black holes.  

\acknowledgments

We would like to thank Michael Nowak for significant
comments and discussion, as well as J\"orn Wilms, Peter Biermann, Tom
Maccarone, and the anonymous referee for suggested improvements.
Support for this work was provided by an NSF Astronomy \& Astrophysics
postdoctoral fellowship, under NSF Award AST-0201597.


\begin{thebibliography}{47}
\expandafter\ifx\csname natexlab\endcsname\relax\def\natexlab#1{#1}\fi

\bibitem[{{Baganoff}(2003)}]{Baganoff2003}
{Baganoff}, F.~K. 2003, AAS/High Energy Astrophysics Division, 35,

\bibitem[{Baganoff {et~al.}(2001)Baganoff, Bautz, Brandt, Chartas, Feigelson,
  Garmire, Maeda, Morris, Ricker, Townsley, \& Walter}]{Baganoffetal2001}
Baganoff, F.~K., Bautz, M.~W., Brandt, W.~N., Chartas, G., Feigelson, E.~D.,
  Garmire, G.~P., Maeda, Y., Morris, M., Ricker, G.~R., Townsley, L.~K., \&
  Walter, F. 2001, Nature, 413, 45

\bibitem[{Baganoff {et~al.}(2003)Baganoff, Maeda, Morris, Bautz, Brandt, \&
  Burrows}]{Baganoffetal2003}
Baganoff, F.~K., Maeda, Y., Morris, M., Bautz, M.~W., Brandt, W.~N., \&
  Burrows, D.~N. 2003, \apj, 591, 891

\bibitem[{{Beloborodov}(1999)}]{Beloborodov1999}
{Beloborodov}, A.~M. 1999, \apjl, 510, L123

\bibitem[{{Bower} {et~al.}(2002{\natexlab{a}}){Bower}, {Falcke}, \&
  {Mellon}}]{BowerFalckeMellon2002}
{Bower}, G.~C., {Falcke}, H., \& {Mellon}, R.~R. 2002{\natexlab{a}}, \apjl,
  578, L103

\bibitem[{{Bower} {et~al.}(2002{\natexlab{b}}){Bower}, {Falcke}, {Sault}, \&
  {Backer}}]{Boweretal2002}
{Bower}, G.~C., {Falcke}, H., {Sault}, R.~J., \& {Backer}, D.~C.
  2002{\natexlab{b}}, \apj, 571, 843

\bibitem[{{Bower} {et~al.}(2003){Bower}, {Wright}, {Falcke}, \&
  {Backer}}]{Boweretal2003}
{Bower}, G.~C., {Wright}, M.~C.~H., {Falcke}, H., \& {Backer}, D.~C. 2003,
  \apj, 588, 331

\bibitem[{{Brunthaler} {et~al.}(2001){Brunthaler}, {Bower}, {Falcke}, \&
  {Mellon}}]{Brunthaleretal2001}
{Brunthaler}, A., {Bower}, G.~C., {Falcke}, H., \& {Mellon}, R.~R. 2001, \apjl,
  560, L123

\bibitem[{{Corbel} {et~al.}(2000){Corbel}, {Fender}, {Tzioumis}, {Nowak},
  {McIntyre}, {Durouchoux}, \& {Sood}}]{Corbeletal2000}
{Corbel}, S., {Fender}, R.~P., {Tzioumis}, A.~K., {Nowak}, M., {McIntyre}, V.,
  {Durouchoux}, P., \& {Sood}, R. 2000, \aap, 359, 251

\bibitem[{{Corbel} {et~al.}(2003){Corbel}, {Nowak}, {Fender}, {Tzioumis}, \&
  Markoff}]{Corbeletal2003}
{Corbel}, S., {Nowak}, M., {Fender}, R.~P., {Tzioumis}, A.~K., \& Markoff, S.
  2003, \aap, {400}, 1007

\bibitem[{Cotera {et~al.}(2004)Cotera, Markoff, Geballe, \& Falcke}]{GC2002}
Cotera, A., Markoff, S., Geballe, T., \& Falcke, H., eds. 2004, Proc.\ 2002
  Galactic Center Workshop, Astron.\ Nachr.\ 324, S1 (Weinheim: Wiley VCH)

\bibitem[{Eckart {et~al.}(2004)Eckart, Baganoff, Morris, Bautz, Brandt,
  Garmire, Genzel, Ott, Ricker, Straubmeier, Viehmann, \&
  Sch\"odel}]{Eckartetal2004}
Eckart, A., Baganoff, F., Morris, M., Bautz, M.~W., Brandt, W., Garmire, G.,
  Genzel, R., Ott, T., Ricker, G., Straubmeier, C., Viehmann, T., \& Sch\"odel,
  R. 2004, \aap, in press (astro-ph/0403577)

\bibitem[{{Eisenhauer} {et~al.}(2003){Eisenhauer}, {Sch{\" o}del}, {Genzel},
  {Ott}, {Tecza}, {Abuter}, {Eckart}, \& {Alexander}}]{Eisenhaueretal2003}
{Eisenhauer}, F., {Sch{\" o}del}, R., {Genzel}, R., {Ott}, T., {Tecza}, M.,
  {Abuter}, R., {Eckart}, A., \& {Alexander}, T. 2003, \apjl, 597, L121

\bibitem[{{Falcke} \& {Biermann}(1995)}]{FalckeBiermann1995}
{Falcke}, H. \& {Biermann}, P.~L. 1995, \aap, 293, 665

\bibitem[{Falcke {et~al.}(2004)Falcke, K\"ording, \&
  Markoff}]{FalckeKoerdingMarkoff2004}
Falcke, H., K\"ording, E., \& Markoff, S. 2004, \aap, 414, 895

\bibitem[{{Falcke} \& {Markoff}(2000)}]{FalckeMarkoff2000}
{Falcke}, H. \& {Markoff}, S. 2000, \aap, 362, 113

\bibitem[{{Falcke} \& {Melia}(1997)}]{FalckeMelia1997}
{Falcke}, H. \& {Melia}, F. 1997, \apj, 479, 740

\bibitem[{{Gallo} {et~al.}(2003){Gallo}, {Fender}, \&
  {Pooley}}]{GalloFenderPooley2003}
{Gallo}, E., {Fender}, R.~P., \& {Pooley}, G.~G. 2003, \mnras, 344, 60

\bibitem[{{Ghez} {et~al.}(2003){Ghez}, {Duch{\^ e}ne}, {Matthews}, {Hornstein},
  {Tanner}, {Larkin}, {Morris}, {Becklin}, {Salim}, {Kremenek}, {Thompson},
  {Soifer}, {Neugebauer}, \& {McLean}}]{Ghezetal2003b}
{Ghez}, A.~M., {Duch{\^ e}ne}, G., {Matthews}, K., {Hornstein}, S.~D.,
  {Tanner}, A., {Larkin}, J., {Morris}, M., {Becklin}, E.~E., {Salim}, S.,
  {Kremenek}, T., {Thompson}, D., {Soifer}, B.~T., {Neugebauer}, G., \&
  {McLean}, I. 2003, \apjl, 586, L127

\bibitem[{{Goldwurm} {et~al.}(2003){Goldwurm}, {Brion}, {Goldoni}, {Ferrando},
  {Daigne}, {Decourchelle}, {Warwick}, \& {Predehl}}]{Goldwurmetal2003}
{Goldwurm}, A., {Brion}, E., {Goldoni}, P., {Ferrando}, P., {Daigne}, F.,
  {Decourchelle}, A., {Warwick}, R.~S., \& {Predehl}, P. 2003, \apj, 584, 751

\bibitem[{{Haardt} \& {Maraschi}(1991)}]{HaardtMaraschi1991}
{Haardt}, F. \& {Maraschi}, L. 1991, \apjl, 380, L51

\bibitem[{{Hannikainen} {et~al.}(1998){Hannikainen}, {Hunstead},
  {Campbell-Wilson}, \& {Sood}}]{Hannikainenetal1998}
{Hannikainen}, D.~C., {Hunstead}, R.~W., {Campbell-Wilson}, D., \& {Sood},
  R.~K. 1998, \aap, 337, 460

\bibitem[{{Heinz} \& {Sunyaev}(2003)}]{HeinzSunyaev2003}
{Heinz}, S. \& {Sunyaev}, R.~A. 2003, \mnras, 343, L59

\bibitem[{{Herrnstein}(1997)}]{Herrnstein1997}
{Herrnstein}, J.~R. 1997, Ph.D.~Thesis

\bibitem[{{Herrnstein} {et~al.}(1999){Herrnstein}, {Moran}, {Greenhill},
  {Diamond}, {Inoue}, {Nakai}, {Miyoshi}, {Henkel}, \&
  {Riess}}]{Herrnsteinetal1999}
{Herrnstein}, J.~R., {Moran}, J.~M., {Greenhill}, L.~J., {Diamond}, P.~J.,
  {Inoue}, M., {Nakai}, N., {Miyoshi}, M., {Henkel}, C., \& {Riess}, A. 1999,
  \nat, 400, 539

\bibitem[{{Ho}(1999)}]{Ho1999}
{Ho}, L.~C. 1999, \apj, 516, 672

\bibitem[{{Liu} {et~al.}(2004){Liu}, {Meyer}, \&
  {Meyer-Hofmeister}}]{LiuMeyerHofmeister2004}
{Liu}, B., {Meyer}, F., \& {Meyer-Hofmeister}, E. 2004, \aap, in press
  (astro-ph/0403460)

\bibitem[{{Liu} \& {Melia}(2002)}]{LiuMelia2002}
{Liu}, S. \& {Melia}, F. 2002, \apjl, 566, L77

\bibitem[{{Markoff} {et~al.}(2001){Markoff}, {Falcke}, {Yuan}, \&
  {Biermann}}]{Markoffetal2001}
{Markoff}, S., {Falcke}, H., {Yuan}, F., \& {Biermann}, P.~L. 2001, \aap, 379,
  L13

\bibitem[{{Markoff} {et~al.}(2003){Markoff}, {Nowak}, {Corbel}, Fender, \&
  Falcke}]{Markoffetal2003}
{Markoff}, S., {Nowak}, M., {Corbel}, S., Fender, R., \& Falcke, H. 2003, \aap,
  397, 645

\bibitem[{{Markoff} \& {Nowak}(2004)}]{MarkoffNowak2004}
{Markoff}, S. \& {Nowak}, M.~A. 2004, \apj, 609, 972

\bibitem[{McClintock \& Remillard(2003)}]{McClintockRemillard2003}
McClintock, J.~E. \& Remillard, R.~A. 2003, in Compact Stellar X-ray Sources,
  Eds. W.H.G. Lewin and M. van der Klis, Cambridge University Press, in press
  (astro-ph/0306213)

\bibitem[{{Meier}(2001)}]{Meier2001}
{Meier}, D.~L. 2001, \apjl, 548, L9

\bibitem[{{Melia} \& {Falcke}(2001)}]{MeliaFalcke2001}
{Melia}, F. \& {Falcke}, H. 2001, \araa, 39, 309

\bibitem[{{Merloni} \& {Fabian}(2002)}]{MerloniFabian2002}
{Merloni}, A. \& {Fabian}, A.~C. 2002, \mnras, 332, 165

\bibitem[{{Merloni} {et~al.}(2003){Merloni}, {Heinz}, \& {di
  Matteo}}]{MerloniHeinzDiMatteo2003}
{Merloni}, A., {Heinz}, S., \& {di Matteo}, T. 2003, \mnras, 345, 1057

\bibitem[{{Nowak} {et~al.}(2004){Nowak}, {Wilms}, {Heinz}, Pooley, \&
  Corbel}]{Nowaketal2004}
{Nowak}, M.~A., {Wilms}, J., {Heinz}, S., Pooley, G., \& Corbel, S. 2004, \apj,
  submitted

\bibitem[{{Page} {et~al.}(2003){Page}, {Breeveld}, {Soria}, {Wu},
  {Branduardi-Raymont}, {Mason}, {Starling}, \& {Zane}}]{Pageetal2003}
{Page}, M.~J., {Breeveld}, A.~A., {Soria}, R., {Wu}, K., {Branduardi-Raymont},
  G., {Mason}, K.~O., {Starling}, R.~L.~C., \& {Zane}, S. 2003, \aap, 400, 145

\bibitem[{{Porquet} {et~al.}(2003){Porquet}, {Predehl}, {Aschenbach}, {Grosso},
  {Goldwurm}, {Goldoni}, {Warwick}, \& {Decourchelle}}]{Porquetetal2003}
{Porquet}, D., {Predehl}, P., {Aschenbach}, B., {Grosso}, N., {Goldwurm}, A.,
  {Goldoni}, P., {Warwick}, R.~S., \& {Decourchelle}, A. 2003, \aap, 407, L17

\bibitem[{{Reid}(1993)}]{Reid1993}
{Reid}, M.~J. 1993, \araa, 31, 345

\bibitem[{{Sch{\" o}del} {et~al.}(2003){Sch{\" o}del}, {Ott}, {Genzel},
  {Eckart}, {Mouawad}, \& {Alexander}}]{Schoedeletal2003}
{Sch{\" o}del}, R., {Ott}, T., {Genzel}, R., {Eckart}, A., {Mouawad}, N., \&
  {Alexander}, T. 2003, \apj, 596, 1015

\bibitem[{{Shakura} \& {Sunyaev}(1973)}]{ShakuraSunyaev1973}
{Shakura}, N.~I. \& {Sunyaev}, R.~A. 1973, \aap, 24, 337

\bibitem[{{Tan} \& {Draine}(2004)}]{TanDraine2004}
{Tan}, J.~C. \& {Draine}, B.~T. 2004, \apj, 606, 296

\bibitem[{{Young} \& {Wilson}(2004)}]{YoungWilson2004}
{Young}, A.~J. \& {Wilson}, A.~S. 2004, \apj, 601, 133

\bibitem[{{Yuan} {et~al.}(2002){Yuan}, {Markoff}, \&
  {Falcke}}]{YuanMarkoffFalcke2002}
{Yuan}, F., {Markoff}, S., \& {Falcke}, H. 2002, \aap, 383, 854

\bibitem[{{Yuan} {et~al.}(2003){Yuan}, {Quataert}, \&
  {Narayan}}]{YuanQuataertNarayan2003}
{Yuan}, F., {Quataert}, E., \& {Narayan}, R. 2003, \apj, 598, 301

\bibitem[{Yuan {et~al.}(2004)Yuan, Quataert, \&
  Narayan}]{YuanQuataertNarayan2004}
Yuan, F., Quataert, E., \& Narayan, R. 2004, \apj, {in press,
  (astro-ph/0401429)}

\end{thebibliography}

\end{document}